\documentclass[
reprint,
superscriptaddress,
 amsmath,amssymb,
 aps,
longbibliography
]{revtex4-2}

\usepackage{physics}

\usepackage{graphicx}% Include figure files
\usepackage{dcolumn}% Align table columns on decimal point
\usepackage{bm}% bold math

\usepackage{xcolor}

\usepackage{mathtools}
\usepackage{gensymb}
\usepackage{lipsum}
\usepackage{comment}
\newcommand*\bfcaption[2]{\caption[#1]{\textbf{#1.}#2}}
\usepackage{xcolor}
\definecolor{UBcolor}{HTML}{007CC1}
\usepackage[colorlinks=true,pdfnewwindow=true,linkcolor=UBcolor,citecolor=UBcolor,urlcolor=UBcolor,breaklinks=true,linktocpage]{hyperref}
\usepackage[all]{hypcap}
\usepackage[nameinlink,capitalise]{cleveref}
\crefname{SI section}{SI Section}{SI Sections}
\Crefname{SI section}{SI Section}{SI Sections}
\usepackage{soul}
\usepackage[normalem]{ulem}

\begin{document}

\preprint{APS/123-QED}

\title{Non-reciprocal torques
 guide self-assembly of active particles\\into clusters with controllable function}

\author{Till Welker}
\email{t.a.welker@sms.ed.ac.uk}
\affiliation{SUPA, School of Physics and Astronomy, University of Edinburgh}
\affiliation{Division of Theoretical Physics, Institute of Physics and Astronomy, Technische Universit\"at Berlin}

\author{Yukino Fujiya}
\affiliation{Division of Theoretical Physics, Institute of Physics and Astronomy, Technische Universit\"at Berlin}

\author{Holger Stark}
\affiliation{Division of Theoretical Physics, Institute of Physics and Astronomy, Technische Universit\"at Berlin}

\begin{abstract}
Self-assembly of constituents determines structure formation in the microscopic world. Attractive forces can assemble active particles into colloidal machines, but they do not fix the particles' orientations, which limits control over the machine's function. We demonstrate that non-reciprocal turn-towards torques not only assemble active particles into clusters, without requiring attractive forces, but also link particle orientations to the cluster configuration. Symmetry then dictates whether the cluster is static, rotates, or translates. In small systems, the particle number uniquely determines the stable configuration and function. In larger systems, there are multiple stable configurations with distinct functions, and tuning the torque strength allows us to bias towards the desired function, such as a run-and-tumble motion. Because the interactions driving assembly can be switched on and off, the clusters self-assemble when needed. For such a ``just-in-time'' self-assembly to be practical, fast assembly is necessary. We show that stochastic resetting, implemented by briefly turning off propulsion and torque, significantly speeds up self-assembly by avoiding slow pathways. Together, our findings demonstrate that non-reciprocal torques can rapidly assemble active particles into colloidal micromachines with controllable function.
\end{abstract}

\maketitle

\section{Introduction} 

To build a macroscopic material or machine, one puts each part in its place. But scaled down, this external manipulation becomes increasingly challenging. An alternative approach is to design the constituents such that they assemble autonomously into the desired target structure. This approach is called \textit{self-assembly}~\cite{whitesides2002self,whitelam2015statistical,zeravcic2017colloquium,nguyen2021organization}, and it regulates diverse processes such as the growth of colloidal crystals~\cite{li2011colloidal}, viruses~\cite{moisant2010exploring,perlmutter2015mechanisms} and cell membranes~\cite{sych2018lipid}, as well as micromachines~\cite{bishop2023active}. There are two key challenges of self-assembly: (1) how can we make the target structure the stable end product, and (2) how can we reach the target structure in a feasible time~\cite{whitelam2015statistical,gartner2022time}.

In passive systems, the ground state of a system can be very well designed using specific ("key-lock")~\cite{sacanna2010lock,hormoz2011design,murugan2015multifarious,zeravcic2017colloquium} or anisotropic~\cite{haxton2013hierarchical,king2024programming} interactions. However, materials built from passive components are passive, such that they are rich in structure but lack dynamic functions. To self-assemble dynamic micro-machines, we require active (e.g. self-propelled) components~\cite{bishop2023active}. The dynamic functions of active clusters are constrained by their symmetry~\cite{brandmuller1986extension,brooks2018shape,aubret2021metamachines}.  Consequently, breaking symmetry allows us to design a specific function; for example, breaking head-tail and time-reversal symmetry enables directed propulsion.

Small stable clusters of active particles, so-called \textit{active colloidal molecules}~\cite{lowen2018active}, provide a rich testing ground. And indeed, one can construct rotors, translators, and chiral swimmers from self-propelled~\cite{gao2013organized,kaiser2015active,guzman2016fission, zhang2016directed, johnson2017dynamic,shen2019hydrodynamic,aubret2021metamachines,subramaniam2024rigid,rosenberg2025windmilling}, nematic~\cite{wykes2016dynamic}, or non-reciprocally interacting~\cite{soto2014self,soto2015self,varma2018clustering,schmidt2019light,subramaniam2024rigid} components. For active molecules built from self-propelled particles, the symmetry of the cluster depends both on the cluster shape and the particle orientations. However, if one only relies on attractive forces to assemble functional clusters, the orientations are randomly distributed during the fabrication process~\cite{johnson2017dynamic,ismagilov2002autonomous}. Thus, when one aims at constructing colloidal machines, this significantly limits the control over symmetry and, consequently, function.  

In this article, we show that non-reciprocal torques turning particle orientations towards their neighbours~\cite{smeets2016emergent,nilsson2017metastable,zhang2021active,das2024flocking,knevzevic2022collective,shea2025emergent,welker2025lattice} promote self-assembly of active-particle clusters without requiring attractive forces. Furthermore, since the symmetry of orientational ordering within the cluster is coupled to the symmetry of the cluster shape, this allows building clusters with highly controllable functions. They range from static to translating, rotating, and even run-and-tumble clusters, when they exhibit multi-stability.

Designing the target state is not enough; we need to ensure that it assembles sufficiently fast. Passive assembly requires finely tailored interactions~\cite{whitelam2015statistical,perlmutter2015mechanisms,king2024programming} and is often slow or results in a low yield. Driving the system out of equilibrium either by active components~\cite{zhu2024proofreading,schubert2025self,dopierala2026odd} or system-wide changes of interaction strength~\cite{faran2025nonequilibrium,sherman2016dynamic,liang2025magnetic} can significantly speed up assembly. However, despite activity, particles can become trapped in ineffective or slow assembly pathways.

We introduce \textit{stochastic resetting}~\cite{evans2011diffusion,evans2020stochastic} as a viable method to expedite self-assembly. We implement resets by briefly turning off activity, which always dissolves the cluster. This cuts off slow assembly pathways, while not changing the target structure. Whether resetting expedites assembly can be predicted from the free assembly-time distribution~\cite{reuveni2016optimal,pal2022inspection}. Resetting, therefore, provides an efficient tool to control assembly, which is easy to implement and predictable.

Altogether, our work establishes a route towards the rapid self-assembly of active clusters, where particle number and non-reciprocal torques determine symmetry and function. 

\vspace{15.5mm}

\section{Results}
\subsection{Active particles with turn-towards torque assemble into dynamic clusters}

\begin{figure}
    \centering
    \includegraphics[scale=0.78]{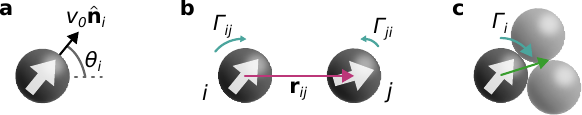}
    \caption{\textbf{Model.} \textbf{a)} Single particle propelling with velocity $v_0 \hat{\mathbf{n}}_i$. \textbf{b)} The turn-towards torque $\Gamma_{ij}$ turns the orientation of particle $i$ towards its neighbour with distance vector $\mathbf{r}_{ij}$. \textbf{c)} 
    Particle $i$ experiences the total torque $\Gamma_i$, orienting it towards the mean distance vector of its neighbours $\langle  \hat{\mathbf{r}}_{ij}  \rangle_{j\in S_i}$ marked by the green arrow.}
    \label{Fig:Model}
\end{figure}

We consider small collectives of active particles in two dimensions. Particles propel with velocity $v_0$ along their orientation vector
$\hat{\mathbf{n}}_i = (\cos\,\theta_i,\sin\,\theta_i)^T$, as shown in Fig.~\ref{Fig:Model}a), and experience steric repulsion, which we implement by a WCA potential. Particles turn towards neighbouring particles through the torque
\begin{equation} \label{eq:torque_begin}
    \mathbf{\Gamma}_{ij}=\mathbf{\Gamma}(\hat{\mathbf{n}}_i,\mathbf{r}_{ij})= \Gamma_0 \, \hat{\mathbf{n}}_i  \times \hat{\mathbf{r}}_{ij} ~~~\text{if}~~~|\mathbf{r}_{ij}|<R\,,
\end{equation}
that particle $i$ experiences in the presence of $j$, as visualised in Fig.~\ref{Fig:Model}b). We choose the cut-off radius $R$ such that only nearest neighbours interact. The particles move in a fluid environment, which overdamps their motion and induces thermal fluctuations in position and orientation, the intensities of which are governed by the translational and rotational diffusivities $D_\mathrm{t}$, $D_\mathrm{r}$, respectively. The Langevin equations of motion, parameters, non-dimensionalisation, and numerical methods are 
described in the methods section~\ref{subsec.methods.active}.

If a particle has multiple neighbours, it orients towards the mean distance vector $\langle  \hat{\mathbf{r}}_{ij}  \rangle_{j\in S_i}$ of the local neighbourhood $S_i$ of particle $i$, as shown in Fig.~\ref{Fig:Model}c). Therefore, the turn-towards torque links the orientation of a particle to its position relative to its neighbours. Particles at the interface of a cluster turn inward, which assembles and stabilises the cluster. Examples of clusters composed of different numbers of particles are shown in Figs.~\ref{Fig:Symmetry}a)-c) and Fig.~\ref{Fig:Free_Particle}a). Similar clustering mechanisms have been described in Refs.~\cite{yan2016reconfiguring,zhang2021active}. 

Once assembled, these clusters behave approximately as rigid bodies, their state is captured by the centre of mass  $\mathbf{r}_\mathrm{c} = \sum_{i=1}^N\mathbf{r}_i/N$  and the cluster orientation. By summing over the single particle dynamics in Eq.~(\ref{eq:positions}), we obtain
\begin{align}\label{Eq:Unit_Velocity}
    \frac{\dd}{\dd t} \mathbf{r}_\mathrm{c} = \frac{1}{N}\sum_{i=1}^N \frac{\dd}{\dd t} \mathbf{r}_i &= v_0 \langle \hat{\mathbf{n}}_i \rangle_\mathrm{c}  + \sqrt{2\frac{D_\mathrm{t}}{N}}\,\boldsymbol{\eta}^\mathrm{t}\,,
\end{align}
with $\boldsymbol{\eta}^\mathrm{t} = \sum_i \boldsymbol{\eta}^\mathrm{t}_i /\sqrt{N}$ being delta-correlated white noise with zero mean and unit variance. The cluster's propulsion velocity $\mathbf{v}_\mathrm{c} = v_0 \langle \hat{\mathbf{n}}_i \rangle_\mathrm{c}$ is dictated by the orientation averaged over the particles in the cluster~\cite{aubret2021metamachines} and the translational diffusivity scales inversely with the cluster size. The steric repulsion is reciprocal and therefore cancels when summing over all particles. 

Likewise, we know that for each point of a rigid body one can always write its velocity as $\dd \mathbf{r}_i /\dd t = \mathbf{v}_\mathrm{c} + \boldsymbol{\omega}_\mathrm{c} \times (\mathbf{r}_i - \mathbf{r}_\mathrm{c})$. So, taking the cross product of $\mathbf{r}_i$ with Eq.\ (\ref{eq:positions}), neglecting noise, and summing over $i$, one ultimately determines the unique angular velocity as
\begin{align}\label{Eq:Unit_AngularVelocity}
    \omega_\mathrm{c} = v_0 \frac{\sum_{i=1}^N (\mathbf{r}_i - \mathbf{r}_\mathrm{c}) \times \hat{\mathbf{n}}_i}{\sum_{i=1}^N 
(\mathbf{r}_i - \mathbf{r}_\mathrm{c})^2} \, .
\end{align}

We quantify translational and rotational
cluster motion from numerical simulations using the centre-of-mass mean squared displacement (MSD) $\langle \Delta \mathbf{r}_\mathrm{c}^2(t) \rangle$ and the mean squared angular displacement (MSAD) $\langle \Delta \theta_i(t)^2\rangle$, respectively. If the orientations are locked, the rotations of the individual particles $\Delta \theta_i$ coincide with the rotation of the rigid cluster. For overdamped particles, ballistic regimes, $\mathrm{MSD}\propto t^2$ and $\mathrm{MSAD}\propto t^2$, indicate directed active translation and rotation, respectively.

\subsection{Symmetry and functionality}~\label{sec:symmetry}
In the previous section we established that turn-towards torques stabilise clusters. We now show that in small systems the cluster's symmetry alone determines which types of directed motion are allowed, while in larger systems there are additional unconstrained degrees of freedom.

\subsubsection{Symmetry dictates function in small clusters}
\begin{figure}
    \centering
    \includegraphics[scale=0.78]{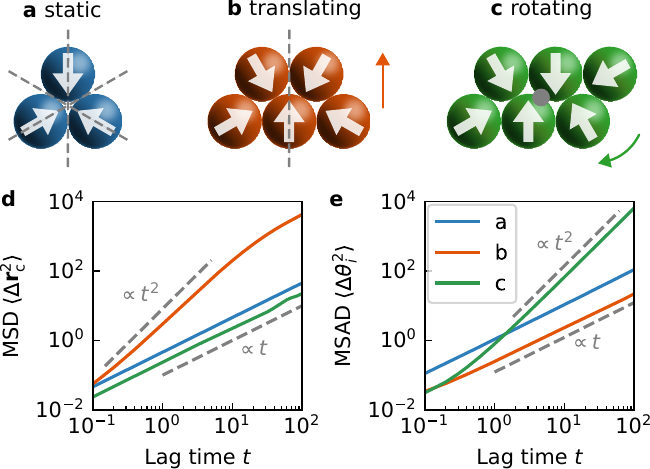}
    \bfcaption{Symmetry and functionality}{ 
    \textbf{a)} Clusters with two or more symmetry axes do not perform directed translation or rotation. \textbf{b)} Clusters with only one symmetry axis, can translate parallel to that axis. \textbf{c)} Clusters with point symmetry cannot translate but rotate. \textbf{d)} Mean squared displacement of the centre of mass $\mathbf{r}_\mathrm{c}$ and,  \textbf{e)} mean squared angular displacement of particle orientations $\theta_i$ for the clusters shown in a-c).
    }
    \label{Fig:Symmetry}
\end{figure}

Turn-towards torques relate symmetries of the positions and orientations to each other. As illustrated in Fig.~\ref{Fig:Model}c), particles turn towards the mean distance vector in their neighbourhood $\langle \hat{\mathbf{r}}_{ij} \rangle_{j\in S_i}$. Consequently, the preferred orientation follows the local symmetry of the particle neighbourhood. If all local symmetries are broken and the torque dominates thermal fluctuations, the total symmetry of the particle orientations reflects the symmetric shape of the cluster, as shown in Fig.~\ref{Fig:Symmetry}a-c). The symmetries in particle positions and orientations then determine the dynamic properties of the cluster.

We first consider a cluster which is mirror-symmetric with respect to one axis. In Fig.~\ref{Fig:Symmetry}b) this is the vertical $y$-axis, such that the cluster is invariant under the transformation $\mathbf{O}=(x_i,y_i,n_i^x,n_i^y)\to (-x_i,y_i,-n_i^x,n_i^y)=\mathbf{O}'$. Inserting in Eq.~(\ref{Eq:Unit_Velocity}) and Eq.~(\ref{Eq:Unit_AngularVelocity}) shows that translational and angular propulsion velocities transform as $\mathbf{P}=(v_\mathrm{c}^x,v_\mathrm{c}^y,\omega_\mathrm{c}) \to (-v_\mathrm{c}^x,v_\mathrm{c}^y,-\omega_\mathrm{c})=\mathbf{P}'$. The Neumann-Minnigerode-Curie principle states that properties $\mathbf{P}$ are at least as symmetric as the object $\mathbf{O}$~\cite{brandmuller1986extension}; in particular, if $\mathbf{O}=\mathbf{O}'$, after a transformation, then $\mathbf{P}=\mathbf{P}'$. This means that clusters with one or more symmetry axes cannot rotate ($\omega_\mathrm{c}=0$) 
or translate perpendicular to the axis ($v_\mathrm{c}^x=0$). Consequently, the axial-symmetric clusters in Fig.~\ref{Fig:Symmetry}a) and b) have purely diffusive angular dynamics as shown by the MSAD in Fig.~\ref{Fig:Symmetry}e). While a symmetry axis suppresses propulsion 
perpendicular to the axis ($v_\mathrm{c}^x =0$), propulsion parallel to the axis ($v_\mathrm{c}^y\neq 0$) is allowed. Indeed, the five particle cluster in Fig.~\ref{Fig:Symmetry}b) has a ballistic regime in the MSD shown in Fig.~\ref{Fig:Symmetry}d). If there are two symmetry axes, propulsion along both directions is suppressed, such that there is no directed motion in two dimensions. An example is sketched in Fig.~\ref{Fig:Symmetry}a) with Fig.~\ref{Fig:Symmetry}d) showing that there is no ballistic regime in the MSD. 

Next, we consider a cluster which is invariant under a rotation by an angle $\alpha$ performed by the rotation matrix $\mathbf{R}_\alpha$. The positions and orientations transform as \mbox{$(\mathbf{r}_i,\hat{\mathbf{n}}_i) \to (\mathbf{R}_\alpha\mathbf{r}_i,\mathbf{R}_\alpha\hat{\mathbf{n}}_i)$}. The translational and angular velocities in Eq.~(\ref{Eq:Unit_Velocity}) and Eq.~(\ref{Eq:Unit_AngularVelocity}) transform as $(\mathbf{v}_\mathrm{c},\omega_\mathrm{c}) \to (\mathbf{R}_\alpha \mathbf{v}_\mathrm{c},\omega_\mathrm{c})$. Now, symmetry demands that the translational propulsion velocity vanishes, while the angular velocity can be non-zero. This is the case for the cluster sketched in Fig.~\ref{Fig:Symmetry}c), for which the MSD is purely diffusive, while the MSAD shows a pronounced ballistic regime, as shown in Fig.~\ref{Fig:Symmetry}d) and e).

The connection of symmetry and function in active colloidal molecules has been previously observed in Refs.~\cite{varma2018clustering,soto2014self,schmidt2019light}. In the absence of any symmetry, the translational and rotational velocities are in general non-zero, which is expected to result in a spiral motion.

\subsubsection{Bulk particles rotate freely}

\begin{figure}
    \centering
    \includegraphics[scale=0.78]{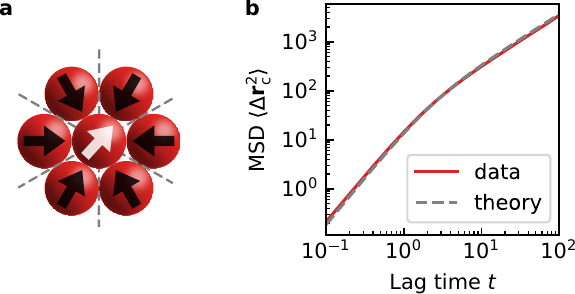}
    \bfcaption{Free inner particle}{ \textbf{a)} Cluster of seven particles. The turn-towards torque orients the outer particles inward. In contrast, the torque on the inner particle vanishes because of the symmetry of the environment, and it can diffuse freely. \textbf{b)} Centre-of-mass mean squared displacement over time obtained from simulation (red solid curve) and predicted assuming a freely diffusing inner particle and perfectly symmetric outer particles (grey dashed curve).
    }
    \label{Fig:Free_Particle}
\end{figure}

The connection between the symmetry of particle positions and symmetry of orientations requires all particle orientations to be linked to the positions. However, if a particle has a symmetric environment (for example in the bulk), the torques acting from the neighbours cancel, and the particle orientation diffuses freely. Let us consider the example sketched in Fig.~\ref{Fig:Free_Particle}a). The outer particles are all oriented towards the centre and their propulsions cancel by symmetry. In contrast, the torques on the inner particle cancel and its orientation diffuses freely. Since its propulsion is not compensated, the cluster performs directed motion.

To model the cluster's motion, let us consider the case of a freely rotating inner particle driving the motion of the cluster, while all other propulsion velocities cancel. Equation (\ref{Eq:Unit_Velocity}) gives the cluster velocity $v_\mathrm{c} = v_0/7$  and translational diffusivity $D_\mathrm{t,c} = D_\mathrm{t}/7 =1/21$. Since only the inner particle contributes to the propulsion of the cluster, the persistence time $\tau_\mathrm{c}$ of the cluster's motion is equal to the persistence time of the inner particle's freely diffusing orientation; thus $\tau_\mathrm{c} =D_\mathrm{r}^{-1}= 1$. Since the cluster behaves as an active Brownian particle, we can immediately formulate the MSD using the known formula~\cite{zottl2016emergent},
\begin{align*}
    \langle \Delta r ^2\rangle (t)= 4D_{\mathrm{t},\mathrm{c}} t + 2v_\mathrm{c}^2\tau_\mathrm{c} t - 2v_\mathrm{c}^2\tau_\mathrm{c}^2(1-e^{-t/\tau_\mathrm{c}}) \,.
\end{align*}
As shown in Fig.~\ref{Fig:Free_Particle}b), the model agrees well with the simulated dynamics of the cluster. On short timescales, we slightly underestimate the propulsion. This could be caused by noise-induced misalignment of the outer particles, leading to an additional contribution to the net propulsion. On long timescales, we slightly overestimate the effective diffusivity $D_\mathrm{eff} = D_{\mathrm{t},\mathrm{c}} + v_0^2\tau_\mathrm{r} /2$~\cite{zottl2016emergent}. This could be caused by the misalignment of the outer particles, introducing an additional noise source to the inner particle's orientation, which reduces its correlation time. Nevertheless, the model of an inner particle with freely diffusing orientation dictating the motion of the cluster explains the observed dynamics quantitatively without any fit parameters.

\subsection{Multi-stability}\label{sec:multistability}

\begin{figure*}
    \centering
    \includegraphics[scale=0.78]{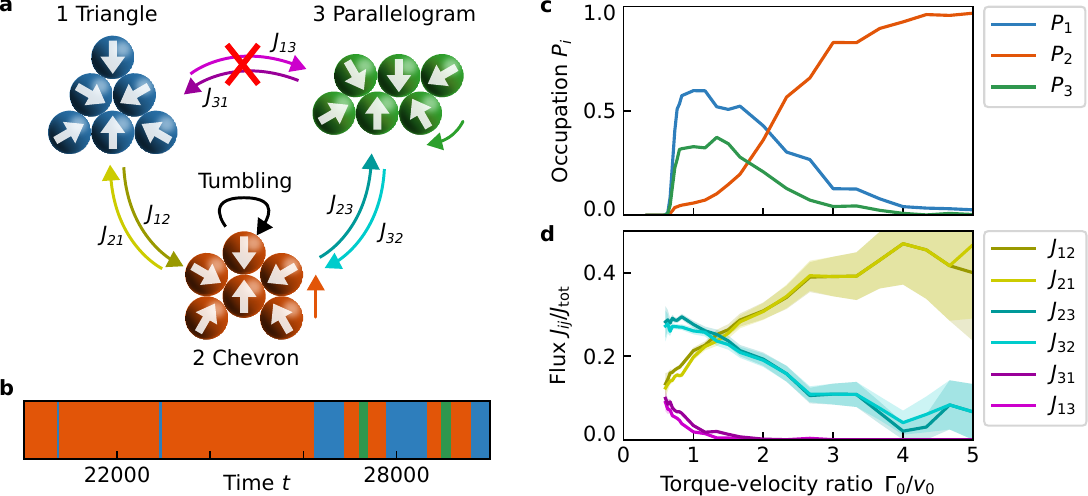}
    \bfcaption{Multi-stability}{ \textbf{a)} Metastable six-particle clusters: static triangle (1, blue), translating chevron (2, orange), and rotating parallelogram (3, green). The arrows show transitions between the states with fluxes $J_{ij}$. The chevron can transition between different arrangements, which results in a tumbling event. \textbf{b)} Time series of a six-particle system switching between the three metastable states (for $\Gamma_0=100$).
    \textbf{c)} Occupation probability $P_i$  to observe the system in state $i$ over torque-velocity ratio $\Gamma_0/v_0$, with $i=1,2,3$ corresponding to triangle, chevron, and parallelogram, respectively. \textbf{d)} Fluxes between the states, $J_{ij}$, also plotted \emph{vs.} $\Gamma_0/v_0$. The fluxes are normalised with the total flux $J_\mathrm{tot}=\sum_{ij}J_{ij}$. The shaded area shows the error (assuming Poissonian transition statistics, $\Delta N_{ij}= \sqrt{\langle N_{ij}\rangle}\approx\sqrt{ N_{ij}}$). For c) and d), we fix $v_0 = 30$ and tune $\Gamma_0$.}
    \label{Fig:Multistability1}
\end{figure*}

So far, we have shown that the functions of the clusters depend on their symmetry. For clusters up to five particles, there is only one stable state for each particle number, such that particle number dictates function. In larger systems, however, multiple cluster shapes are metastable. We discuss here the six-particle system, as it is the smallest multi-stable one and rich in dynamical function.

In the six-particle system, we observe three metastable states shown in Fig.~\ref{Fig:Multistability1}a): triangle, chevron, and parallelogram. Again, symmetry dictates functionality, such that the triangle is stationary, the chevron translates, and the parallelogram rotates~\cite{varma2018clustering}. By controlling the occupation probabilities of the metastable states, we could therefore tune the cluster's function. 

We use the algorithm described in Appendix~\ref{sec:cluster_detection} to measure the time series of the cluster's state shown in Fig.~\ref{Fig:Multistability1}b). In the following, we will show that the occupation probabilities of the metastable states and also the transition probabilities or fluxes between the states depend on the torque-velocity ratio. There are frequent transitions between states analogous to transitions between mesostates in a free energy landscape. Additionally, there are transitions between different arrangements ("microstates") of the chevron, which result in abrupt reorientations (tumbling).

\subsubsection{Occupation Probabilities and Reconfigurations}
The occupation probability of the system being in the triangle ($i=1$), chevron ($i=2$), or parallelogram ($i=3$) state depends on the ratio of torque and velocity $\Gamma_0/v_0$ as shown in Fig.~\ref{Fig:Multistability1}c). For very small torques, there are no clusters. But as the torque-velocity ratio surpasses $\Gamma_0/v_0\approx2/3$, triangle and parallelogram clusters form. Increasing $\Gamma_0/v_0$ increases the probability of a chevron cluster until it completely dominates the population around $\Gamma_0/v_0\sim 5$. 

To understand the onset of clustering, we consider the relevant timescales in our system: (1) the time to propel over the interaction range, $\tau_v = R/v_0$, and (2) the time for the torque to reorient the particle, $\tau_\Gamma =\gamma_\mathrm{r}/\Gamma_0$. For clusters to form, particles need to reorient in the time they pass each other, such that we expect an onset of clustering at $\tau_v \sim \tau_\Gamma$. Inserting the definitions of the timescales and rearranging yields $\Gamma_0/v_0 \sim \gamma_\mathrm{r}/R = 2/3$, which quantitatively agrees with the onset of clustering in Fig.~\ref{Fig:Multistability1}c).

For torques above the threshold of cluster formation, we observe triangle, chevron, and parallelogram clusters. The occupation probabilities $P_i$ can be tuned by the torque-velocity ratio, as shown in Fig.~\ref{Fig:Multistability1}c). For equilibrium systems, one can predict $P_i$ from energetic and entropy arguments, as it was done for passive attractive colloids in Ref.~\cite{perry2015two}. But our clusters are stabilised by non-equilibrium propulsion forces and non-reciprocal torques; there is no free energy, and transition kinetics becomes relevant. Active motion can suppress or enhance fluctuations. While the propulsion velocity sets the strength of this effect, the torque determines how fast particles reorient as a response to the perturbed neighbourhood. The interplay between these two factors results in a non-trivial dependence of the occupation probabilities on the torque-velocity ratio. Therefore, we can use the torque strength to control the system and bias it towards desired structures and, consequently, functions.

We quantify the transitions between different cluster shapes, by measuring the flux from state $i$ to $j$, $J_{ij} = \langle \mathrm{d} {N}_{ij}/\mathrm{d}t\rangle$, with number of transitions $N_{ij}$. All six fluxes are shown in Fig.~\ref{Fig:Multistability1}d). Notably, for $\Gamma_0/v_0 >1$ transitions between triangle and parallelogram are almost completely suppressed $J_{31},J_{13}\approx 0$, even if the chevron cluster has low occupation probability. As a consequence, the network of states is effectively linear (triangle $\leftrightharpoons$ chevron $\leftrightharpoons$  parallelogram). In such linear networks, there cannot be a cyclic current, and fluxes must balance in the steady state, $J_{ij}\approx J_{ji}$, in agreement with Fig.~\ref{Fig:Multistability1}d). This makes the configuration space equilibrium-like, with transitions obeying effectively detailed balance. This is noteworthy because detailed balance is generally broken for assembly cycles out of equilibrium~\cite{bishop2023active}.

\subsubsection{Run-and-Tumble motion}

\begin{figure}
    \centering
    \includegraphics[scale=0.78]{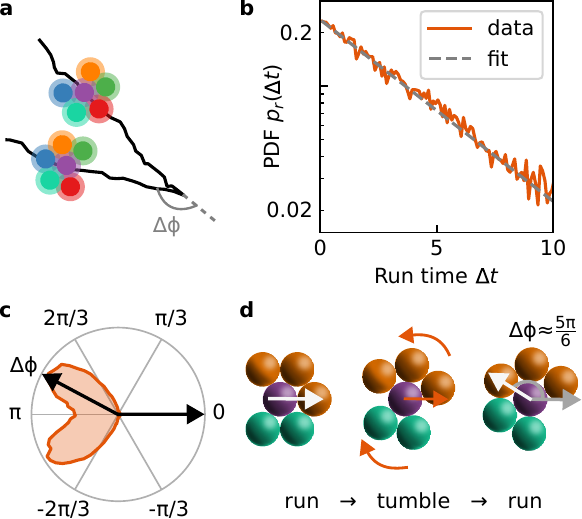}
    \bfcaption{Tumbling}{ \textbf{a)} The chevron cluster exhibits tumbling events in which the leader particle changes. This reorients the propulsion direction by $\Delta \phi$. \textbf{b)}  Distribution $p_\mathrm{r}(\Delta t)$ of run times $\Delta t$. It can be well fitted by the exponential $p_\mathrm{r,fit}(\Delta t)=ce^{-ct}$ with tumbling rate $c= 0.235$.
    \textbf{c)} Probability distribution $p_\phi(\Delta\phi)$ of tumbling angle $\Delta \phi$ and  \textbf{d)} typical tumbling event with the central particle pushing forward and two and three outer particles rotating around it.
    }
    \label{Fig:Tumbling}
\end{figure}

The six-particle system has three different metastable morphologies ('mesostates'). Because all particles are identical, there are multiple sub-configurations ('microstates') resulting in the same mesostate. And indeed, for the translating chevron cluster we observe transitions between those microstates. These reconfigurations change
the ``{}leader particle" and consequently a rapid reorientation
of the propulsion direction $\hat{\mathbf{v}}_\mathrm{c} = \mathbf{v}_\mathrm{c}/|\mathbf{v}_\mathrm{c}|$, with cluster velocity $\mathbf{v}_\mathrm{c}$, occurs,
as shown in Fig.~\ref{Fig:Tumbling}a). The reorientations resemble tumble events resulting in a run-and-tumble motion of the chevron cluster. The run-and-tumble motion within one mesostate sets the dynamics apart from previously reported run-and-tumbling in active colloidal molecules, where tumbling is caused by transitions between a static and translating state~\cite{soto2015self,wykes2016dynamic}.

We identify tumbling events by registering transitions between two chevron clusters as discussed in Appendix~\ref{sec:cluster_detection}. From simulations, we measure the distribution of run times between tumbling events $p_\mathrm{r}(\Delta t)$ shown in Fig.~\ref{Fig:Tumbling}b). It is well fitted by an exponential $p_\mathrm{r,fit}(\Delta t)=ce^{-ct}$ with tumbling rate $c = 0.235$.
The exponential run time distribution agrees with the notion of a barrier crossing with a constant transition rate. This further supports the picture of reconfigurations as transitions between micro- and mesostates.

During each tumbling event, the cluster reorients. We numerically measure the angular change $\Delta \phi$ of the propulsion direction $\hat{\mathbf{v}}_\mathrm{c}$ during the tumbling events. The resulting tumbling angle distribution $p_\mathrm{t}(\Delta \phi)$ is shown Fig.~\ref{Fig:Tumbling}c). It is highly non-uniform, with two symmetric peaks corresponding to an almost full inversion of the propulsion direction. The peaks are associated with the reconfiguration shown in Fig.~\ref{Fig:Tumbling}d), and its chiral counterpart. Thermal fluctuations break the bond between the leader particle and one of its neighbours. Now the central particle can push forward, and the two and three connected outer particles move along the central particle and recombine on the other side. 
If these two groups of particles rotate by approximately the same angle, the tumbling angle is $\pm 5\pi/6$, which agrees with the two peaks of the tumbling distribution in Fig.~\ref{Fig:Tumbling}c).

We now show that the tumbling events contribute significantly to the decorrelation of the propulsion direction. Because our run times are exponentially distributed, and tumbling events are almost instantaneous, the cluster orientation correlation function is expected to be~\cite{datta2024random}
\begin{align*}\label{eq:velocity_correlation}
    \langle \hat{\mathbf{v}}_\mathrm{c}(t)\cdot \hat{\mathbf{v}}_\mathrm{c}(0)\rangle = e^{-kt} \text{ with } k = D_{\mathrm{r,c}} + c(1-\langle \cos \Delta\phi\rangle_{p_\mathrm{t}}) \,.
\end{align*}
The total decorrelation rate $k$ combines the cluster's orientational diffusion, $D_{\mathrm{r,c}}$, and the decorrelation due to tumbling,
$1-\langle \cos \Delta\phi\rangle_{p_\mathrm{t}}$, where $c$ is the tumbling rate introduced before. For the torque $\Gamma_0 = 100$, the mean reorientation after a tumbling event is $\langle \cos \Delta\phi\rangle_{p_\mathrm{t}} = -0.721$. The negative value reflects the tendency to revert the orientation. Comparing the decorrelation contribution of tumbling, $c(1-\langle \cos \Delta\phi\rangle_{p_\mathrm{t}}) = 0.404$, to the numerically measured decorrelation rate $k = 0.502$ shows that tumbling has a significant impact on the dynamics of the chevron cluster, and is responsible for most of the decorrelation of the cluster's orientation. This is similar to tumbling \textit{E. coli} bacteria~\cite{berg1972chemotaxis}.

\subsection{Cluster assembly} \label{sec:assembly} 

\begin{figure*}
    \centering
    \includegraphics[scale=0.76]{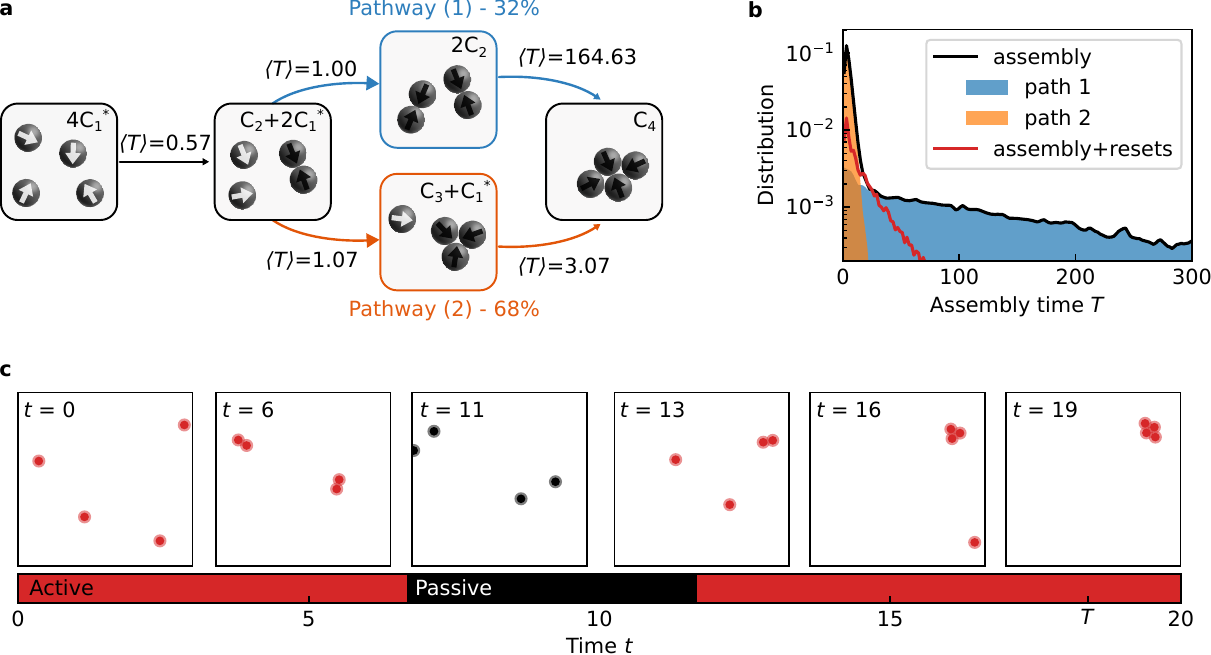}
    \bfcaption{Self-Assembly}{ \textbf{a)} The system starts from four individual particles. Particles collide and merge, resulting in transitions between mesostates. Propelled/stationary clusters are visualised with white/black arrows. Clusters are labelled as $l\mathrm{C}_m$ with $m$ referring to the number of particles in the cluster and $l$ to the number of clusters with that size, superscripts $*$ indicate propelled components. There are two possible assembly pathways (blue and orange), both leading to the final four-particle cluster. The probability of the pathways and the mean waiting time $\langle T\rangle$ of the transitions between mesostates are measured in 65\,000 runs. \textbf{b)} Distribution of assembly times $T$. First, distribution $p_\mathrm{a}$ of the original assembly process (black) with contributions from pathways (1) and (2) in blue and orange, respectively. Second, distribution $p_{\mathrm{a}+r}$ of the assembly process with Poissonian resetting at rate $r=0.1$ and reset duration $\tau = 5$. \textbf{c)} Resetting is implemented by turning off propulsion and torques for a duration of $\tau$. In the passive phase, diffusion redistributes the particles. The reset avoids getting stuck in the slow assembly pathway (1).}
    \label{Fig:Assembly}
\end{figure*}

So far, we have shown that turn-towards torques assemble self-propelled particles into stable clusters. In this section, we study the assembly dynamics. We focus on the four-particle system, because it is the simplest non-trivial case. It has only one final structure, but can self-assemble along two paths, as shown in Fig.~\ref{Fig:Assembly}a). We run 65\,000 simulations in a periodic box of size $20\times20$, starting with one particle randomly placed in each quadrant and terminating when the final structure is reached. The clusters are detected as explained in Appendix~\ref{sec:cluster_detection}. We explore the assembly
pathways and analyse their probabilities and durations using combinatorial and dynamical arguments. Finally, we demonstrate that stochastic resetting can expedite assembly.

\subsubsection{Assembly pathways and timescales}
Let us follow the assembly pathways shown in Fig.~\ref{Fig:Assembly}a). Starting from four free particles, the only possible assembly event is the collision and merger of two particles, resulting in one two-particle cluster and two remaining free particles. But now there are two possible paths to proceed. (1) The remaining free particles form a second two-particle cluster, and afterwards the two clusters merge. (2) One of the free particles merges with the two-particle cluster, and afterwards the resulting three-particle cluster merges with the free particle. Both pathways result in the same final four-particle cluster. In ``{}chemical notation" the pathways can be written as 
\begin{align*}
    &\text{pathway 1:}~~4\mathrm{C}_1^* \to \mathrm{C}_2+2\mathrm{C}_1^*\to 2\mathrm{C}_2 \to \mathrm{C}_4  \text{ and}\\
    &\text{pathway 2:}~~4\mathrm{C}_1^* \to \mathrm{C}_2+2\mathrm{C}_1^*\to \mathrm{C}_3+\mathrm{C}_1^* \to \mathrm{C}_4\,,
\end{align*}
with $l\mathrm{C}_m$ referring to $l$ colloidal molecules formed from $m$ colloids with $*$ indicating self-propelled components.

The pathways (1) and (2) are observed in $32\%$ and $68\%$ of the runs, respectively. To understand why, let us use combinatorics. There is only one way the two free particles can meet to form a second cluster ($\mathrm{C}_2+2\mathrm{C}_1^*\to 2\mathrm{C}_2$), while there are two free particles which can collide with the already existing two-particle cluster to form a three-particle cluster ($\mathrm{C}_2+2\mathrm{C}_1^*\to \mathrm{C}_3+\mathrm{C}_1^*$). This simple argument 
yields a ratio of 1 to 2, which is remarkably consistent with the simulations.

Let us now explore the distribution $p_\mathrm{a}(T)$ of assembly time $T$ shown in Fig.~\ref{Fig:Assembly}b). The mean assembly time is $\langle T \rangle=56.54$. However, the mean alone is not enough to identify the relevant timescales, because $p_\mathrm{a}(T)$ is highly non-Poissonian, with a strong probability weight on very fast assembly and a pronounced tail corresponding to very slow assembly. We decompose the distribution into the contributions $p^{(1)}_\mathrm{a}$ and $p^{(2)}_\mathrm{a}$ from pathway 1 and 2, respectively. The decomposition reveals that assembly along path 1 is typically much slower than assembly along path 2, with mean assembly times $\langle T\rangle_1=166.19$, $\langle T\rangle_2=4.72$, respectively. Resolving the mean assembly time of each step (Fig.~\ref{Fig:Assembly}a) shows that the bottleneck of the assembly is the merger of two dimers into the final structure $2\mathrm{C}_2 \to \mathrm{C}_4$. To understand why, note that the individual particles $\mathrm{C}_1^*$ are self-propelled, while the two $\mathrm{C}_2$ and three $\mathrm{C}_3$ particle clusters are stationary due to their symmetry (see Sec.~\ref{sec:symmetry}). Self-propelled units explore the space more efficiently, resulting in fast encounters, while stationary clusters only diffuse, resulting in fewer encounters. Assembly path 2 has propelled units throughout the assembly process, enabling fast encounters and mergers. For assembly path 1, on the other hand, the step $2\mathrm{C}_2 \to \mathrm{C}_4$ has no propelled units, which significantly slows down the assembly. A similar effect has been reported for active dimers, where translating clusters meet and merge faster than non-motile and rotating clusters~\cite{johnson2017dynamic}.

\subsubsection{Expedite self-assembly with stochastic resetting}

Kinetic traps pose a major challenge for efficient self-assembly. In our system, pathway (1) significantly slows down the assembly process. In this section, we will show how a physical implementation of stochastic resetting can significantly expedite assembly by escaping the slow pathway.

The assembly of the final cluster corresponds to a first passage of the assembled state in state space. Consequently, $p_\mathrm{a}(T)$ can be seen as a first-passage-time distribution. One prominent strategy to speed up first passage times is stochastic resetting~\cite{evans2011diffusion,evans2020stochastic}. The idea is to reset the process with rate $r$ to its initial state to cut off long trajectories. Resetting can reduce the mean first passage time if the standard deviation $\sigma_T$ of $p_\mathrm{a}(T)$ is larger than its mean $\langle T \rangle$ ~\cite{reuveni2016optimal}. This is the case for our system with $\langle T \rangle = 56.54$ and $\sigma_T = 124.56$.

One way to physically implement these ``{}resets" is to turn off propulsion and torque (e.g. by switching off the driving field) for a duration 
$\tau$. During that time, the clusters dissolve and diffusion redistributes the particles. Since small $\tau$ incompletely decorrelates the particles, 
while large $\tau$ delays the assembly, the resets are partial and require finite time. Consequently, the criterion $\sigma_T >\langle T\rangle$ 
alone does not guarantee a benefit.

As a proof of concept, we study Poissonian resetting at rate $r = 0.1$ with reset duration $\tau = 5$ (10\,000 runs). The reset rate is chosen such that the mean time in between resets $r^{-1}=10$ is long enough to avoid strong interference with the fast assembly path, while being short enough to interrupt long trajectories early on. The reset duration $\tau=5$ is large enough for the particles to reorient and diffuse out of the torque interaction range, while still being significantly below the average assembly time of the original process. Figure~\ref{Fig:Assembly}c) shows a system originally following the slow pathway (1). Turning off the activity dissolved the clusters and allows the system to reassemble this time following the fast pathway (2), which is completed before the next reset.
Figure~\ref{Fig:Assembly}b) shows that the resets successfully cut off long assembly trajectories and the mean assembly time is significantly reduced to $\langle T\rangle_r = 16.06$ (speed-up by a factor $3.5$). Additionally, the standard deviation is also significantly reduced $\sigma_{T,r} = 18.62$ (by a factor $6.7$), making the process more predictable. Even without optimising over $r$ and $\tau$, resetting significantly expedites the assembly in this system; this highlights the promises of stochastic resetting for more general self-assembly problems~\cite{faran2025nonequilibrium}.

\section{Discussion}

In this article, we showed that non-reciprocal turn-towards torques assemble active particles into clusters while simultaneously making the 
cluster's function controllable. The turn-towards torques couple the particle orientations to the assembled particle configuration. Therefore, the cluster shape determines the symmetric arrangement of particle orientations and, consequently, whether clusters are static, translating, or rotating. For small clusters, the particle number selects the stable configuration and thereby provides a simple means of programming cluster function. Larger clusters become multistable; the six-particle cluster has a static, translating, and rotating state, and even mimics run-and-tumble motion.
Tuning the torque strength then biases the system towards a desired configuration and function. Particle number and torque therefore provide two complementary means of control.

In this article, we specifically focus on turn-towards torques. But the connection between cluster configuration, symmetry, and function applies more generally to active systems where particle orientations are linked to their neighbours' positions. This principle could be explored experimentally for Janus particles in an AC electric field, which exhibit turn-towards torques~\cite{zhang2021active}, but also for self-aligning particles, for which torques implicitly depend on interaction forces~\cite{baconnier2022selective,baconnier2025self}. Thus, such experiments might further explore and establish the coupling of orientation and cluster configuration as a general route towards controlling the function of active machines.

Both the assembly and function of the clusters are driven by non-equilibrium forces. By tuning activity, clusters can be assembled when 
needed, perform their task, and disassemble afterwards. For this ``just-in-time'' assembly to be practical, fast assembly is crucial. Therefore, slow assembly pathways need to be avoided.
By stochastically turning off activity, we disassemble the intermediate clusters. With these stochastic resets, we avoid getting stuck in kinetic traps and significantly expedite self-assembly, highlighting the utility of controlled \textit{self-disassembly}~\cite{kruse2024active}. 
Future theoretical work should explore how finite reset durations~\cite{reuveni2016optimal,olsen2024thermodynamic}, the remaining correlations after the resets~\cite{tal2022diffusion}, and feedback control~\cite{faran2025nonequilibrium} affect the efficiency of assembly. At the same time, experiments could test this principle in colloidal systems with metastable intermediate states. Theory and experiment could establish controlled disassembly using stochastic resetting as a versatile tool for a broad range of non-equilibrium assembly processes.

Our work demonstrates how non-reciprocal torques guide the self-assembly of active particles into small clusters, with functions that can be controlled by particle number and torque strength. With stochastic resetting, it also suggests a means to speed up self-assembly. Thus, our article might serve as an inspiration for experimental and further theoretical work on building functional materials and machines on the 
microscale through self-assembly.

\section{Methods}
\subsection{Active Particles with non-reciprocal torques}
\label{subsec.methods.active}

We consider small collectives of particles in two dimensions. Each particle propels with velocity $v_0$ along its orientation $\hat{\mathbf{n}}_i = (\cos\,\theta_i,\sin\,\theta_i)^T$ as shown in Fig.~\ref{Fig:Model}a). Steric repulsion avoids strong particle overlap. The repulsion of two particles with distance $\mathbf{r}_{ij}=\mathbf{r}_{j}-\mathbf{r}_{i} = |\mathbf{r}_{ij}| \hat{\mathbf{r}}_{ij}$ is modeled by the WCA potential $U(\mathbf{r}_{ij})$, which is $U(\mathbf{r}_{ij}) = 4\epsilon [ (\sigma/|\mathbf{r}_{ij}|)^{12}-(\sigma/|\mathbf{r}_{ij}|)^{6}] + \epsilon$ for $|\mathbf{r}_{ij}|<2^{1/6}\sigma$ and $U(\mathbf{r}_{ij})=0$ otherwise. The parameter $\sigma$ sets the particle diameter. Particles turn-towards neighbours within a cutoff radius $R$, which we describe by the torque
\begin{equation*}
    \mathbf{\Gamma}_{ij}=\mathbf{\Gamma}(\hat{\mathbf{n}}_i,\mathbf{r}_{ij})= \Gamma_0 \, \hat{\mathbf{n}}_i  \times \hat{\mathbf{r}}_{ij} ~~~\text{if}~~~|\mathbf{r}_{ij}|<R\,,
\end{equation*}
experienced by particle $i$ in the presence of $j$, as visualised in Fig.~\ref{Fig:Model}b). The particles move in a fluid environment, which overdamps their motion and also induces thermal fluctuations. Altogether, the dynamics of particle $i$ with position $\mathbf{r}_i$ and orientation $\theta_i$ reads
\begin{align}
    \frac{\dd}{\dd t} \mathbf{r}_i &= v_0 \hat{\mathbf{n}}_i - 
    \frac{1}{\gamma_\mathrm{t}}
     \nabla_i \sum_{j\neq i} U(\mathbf{r}_{ij}) + \sqrt{2D_\text{t}}\,\boldsymbol{\eta}_i^\mathrm{t}, \label{eq:positions}\\
    \frac{\dd}{\dd t} \theta_i &= \frac{1}{\gamma_\mathrm{r}}\sum_{j\neq i} \Gamma_{ij} + \sqrt{2D_\text{r}}\,\eta_i^\mathrm{r}. \label{eq:orientations}
\end{align}
Here, $\eta_i^\alpha$ is delta-correlated white noise with zero mean and unit variance: 
$\langle\eta_i^\alpha(t)\eta_j^\beta(t')\rangle = \delta_{ij}\delta_{\alpha\beta}\delta(t-t')$. For spherical particles the translational diffusivity 
$D_\mathrm{t}$ and rotational diffusivity $D_\mathrm{r}$ are connected by $D_\mathrm{t} = \sigma^2 D_\mathrm{r}/3$, and the Einstein relation, $D_\alpha = k_B T / \gamma_\alpha$, links Stokes friction coefficients and diffusivities.

We non-dimensionalise the system by rescaling lengths by $\sigma$, time by $D_\mathrm{r}^{-1}$, and energy by $k_B T$. The latter is also used to rescale torques. In this paper, we focus on the weak noise regime. Consequently, we choose a large torque amplitude, $\Gamma_0 \gg 1$, and a large propulsion speed  compared to translational diffusion, $v_0=30$ (or Péclet number $\mathrm{Pe} = \sigma v_0/D_\mathrm{t} = 90\gg 1$). If not stated otherwise, we take $\Gamma_0 = 100$ in our simulations. We choose for the strength of the WCA potential $\epsilon = 100$, which guarantees minimal overlap. The range of the orientational interaction is set to $R=1.5\,\sigma$, which is small enough such that only nearest neighbours interact; but large enough such that thermal fluctuations do not move nearest neighbours in and out of the interaction range. We perform simulations of one to seven particles using the Euler method with time step $\Delta t = 10^{-5}$ and a total simulation time $t_\mathrm{tot}=10^5$.

\subsection{Cluster detection} \label{sec:cluster_detection}

\begin{figure}[b]
    \centering
    \includegraphics[scale=0.78]{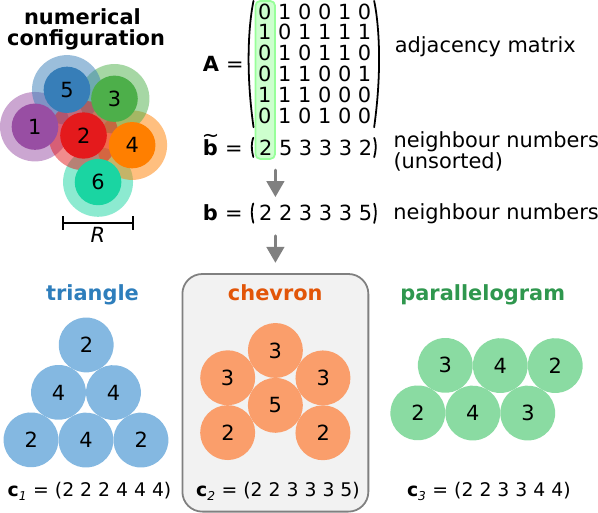}
    \bfcaption{Cluster Analysis}{ From numerical data, we compute the adjacency matrix $\mathbf{A}$, from which we determine the number of neighbours for each particle $\tilde {\mathbf{b}}$. The sorted neighbour numbers $\mathbf{b}$ are compared to the set of stable configurations $\{\mathbf{c}_l\}$ to determine the configuration $l$.}
    \label{Fig:Cluster_Analysis}
\end{figure}

In this section, we describe how to detect the cluster configuration and transitions from numerical data $\mathbf{r}_1(t_i),...,\mathbf{r}_N(t_i)$ with discrete saving times $t_i$. Figure~\ref{Fig:Cluster_Analysis} visualises the procedure.

The adjacency matrix
\begin{align*}
    A_{nm} = 
    \begin{cases} 
        1 & \text{for } |\mathbf{r}_{nm}|\leq R \text{ and } n\neq m \\
        0 & \text{otherwise}
    \end{cases}
\end{align*}
detects if two particles are within the torque interaction range $R$. From the adjacency matrix, we compute the unsorted number of neighbours
\begin{align*}
    \tilde b_n = \sum_m A_{nm} \,.
\end{align*}
This corresponding sorted list
\begin{align*}
    b_n = \mathrm{sort} ( \tilde b_n)
\end{align*}
is ordered such that $b_1\leq b_2 ...\leq b_N$. The sorted neighbour list $\mathbf{b}$ is invariant under particle permutations, and in the cases considered here, the sorted neighbour list uniquely determines the cluster configuration. By comparing $\mathbf{b}$ to the set of stable cluster configurations $C = \{\mathbf{c}_l\}$, we identify the configuration at each time step. This gives us a trajectory of observed cluster configurations $l(t_i)$. During transitions, the cluster may not correspond to any stable configuration ($\mathbf{b}(t_i) \notin C$), in which case we set $l(t_i)=-1$.

To detect transitions between different states, we register instances for which the configuration changes $l(t_i) \neq l(t_{i+1})$. The endpoint of the transition is the next stable configurations ($l(t_j)\neq -1$). A trajectory $(...,3,-1,-1,1,...)$ is a transition from configuration $3$ to configuration $1$, the intermediate states $-1$ are disregarded. Due to the discrete saving times, the intermediate states are not necessarily detected, resulting in trajectories like $(...,3,2,...)$.

Because we do not necessarily resolve intermediate states, the strategy above is not sufficient to reliably resolve transitions between two sub-configurations of the same configuration. Instead, we register changes in the unsorted neighbour list $\tilde {\mathbf{b}}(t_i)\neq \tilde {\mathbf{b}}(t_{i+1})$, which is expected to change when transitioning between different sub-configurations. For the resulting event to be a transition between sub-configurations, two conditions need to be satisfied: (1) The next stable cluster needs to have the same configuration (same sorted neighbour list), but (2) the next stable cluster has a different unsorted neighbour number list. Condition (2) ensures that we do not consider fluctuations which come back to the same sub-configuration. As an example consider $l_{1,2,3} = 2,-1,2$ with $\tilde{\mathbf{b}}(t_{1,2,3})$, this is only a transition between sub-configurations if $\tilde{\mathbf{b}}(t_{1})\neq \tilde{\mathbf{b}}(t_{3})$, otherwise it is just a fluctuation around one sub-configuration.

\bibliography{Bibliography}

\bigskip\noindent\textbf{Data availability -} The code required to reproduce the results of this paper is openly available at \href{https://github.com/tillwelker/Non-reciprocal-torques-guide-self-assembly-of-active-particles}{https://github.com/tillwelker/Non-reciprocal-torques-guide-self-assembly-of-active-particles}.

\bigskip\noindent\textbf{Acknowledgements -} We thank  Juri Schubert, Patrick Pietzonka, Kristian S. Olsen, and Tyler Shendruk for helpful insights into the cluster dynamics and Francesco Mottes, Qian-Ze Zhu, Martin Evans, and Wilson C. K. Poon for their perspectives on resetting and self-assembly. T.W. thanks the University of Edinburgh for their PhD studentship and the Technical University of Berlin for funding a student assistant position.

\bigskip\noindent\textbf{Author contributions -} T.W. and H.S. designed the study; T.W. and Y.F. performed the simulations and analysed the data; T.W. and H.S. wrote the manuscript.

\bigskip\noindent\textbf{Competing interests -} The authors declare no competing interests.

\end{document}